\documentclass[12pt,preprint]{aastex}
\usepackage{verbatim}
\shorttitle{GCs in Simulations}
\shortauthors{Boley et al.}

\begin{document}
\title{Globular Cluster Formation within a Cosmological Context}

\author{Aaron C.\ Boley,  George Lake, Justin Read, \& Romain Teyssier}

\affil{Institute for Theoretical Physics, University of Zurich, Winterthurerstrasse 190,
Zurich, CH-8057, Switzerland; aaron.boley@gmail.com}

\begin{abstract}

We place constraints on the formation redshifts for blue globular clusters (BGCs), independent of the details of hydrodynamics and population III star formation.
The observed radial distribution of BGCs in the Milky Way Galaxy suggests that they formed in biased dark matter halos at high redshift.  As a result, simulations of a $\sim 1$ Mpc box up to $z\sim 10$ must resolve BGC formation in $\Lambda$CDM.  We find that most halo stars could be produced from destroyed BGCs and other low-mass clusters that formed at high redshift.  We present a proof-of-concept simulation that captures the formation of globular-like star clusters.
 \end{abstract}

\keywords{early universe -- globular clusters: general  --  galaxies: star clusters}

\section{Introduction}

The first stars and star clusters are responsible for the initial chemical enrichment of the universe.  They are signatures of the earliest stages of galaxy formation, and their remnants may provide the seeds for supermassive black holes.  The oldest known star clusters are globular clusters (GCs).  In the Milky Way, their color distribution is bimodal (Zinn 1985; Harris 2009), possibly indicating two formation channels.  The blue, metal-poor GCs ($\rm [Fe/H] <-1$) have a typical $\rm [Fe/H]\sim-1.5$ and the red, metal-richer ($\rm [Fe/H] > -1]$) population has a typical $\rm [Fe/H]\sim -0.5$.  Blue GCs (BGCs) have chemical signatures and radial distributions that are similar to halo stars (e.g., Helmi 2008), while the red population is associated with the galactic disk and shows a clear circular velocity component. Relative age estimates show that red GCs are younger than BGCs, with an average age separation of about 1.5 Gyr between the Galactic populations (e.g., de Angeli et al.~2005).  The typical GC in the Galaxy has an $M_V\sim-7.3$ (Harris 1991) and mass $M=1.4\times10^{5} M_{\odot}$, assuming a mass-to-light ratio $\Upsilon_V=2$.  The dark matter content of these systems is small, if present at all (e.g., Moore 1996; Baumgardt et al.~2009).

The formation of GCs remains poorly understood (see Brodie \& Strader [2006] for a summary). 
Motivated by their extreme ages, Peebles \& Dicke (1968) argued for their formation shortly after recombination, preceding protogalaxy formation.  
The GCs are centrally concentrated in the galaxy with half of them closer to the galactic center than the Sun.  In contrast, the
half mass radius of the dark halo is over 10 times larger.  As a result, most subsequent theoretical work has focused on producing them after the galaxy collapses.  
These ideas have included compression behind strong shocks
(Gunn 1980) or thermal instabilities is the early protogalaxy  (Fall \& Rees 1985).   
However, these conditions could never have existed in dwarf galaxies, yet some contain BGCs (e.g., Fornax and WLM; Harris 1991).
So while we reject these ideas here, they can be revived to produce the younger population of GCs in mergers (Ashman \& Zepf 2001). 

BGC ages remain a compelling case for formation before the collapse of the galaxy (e.g., Sarajedini et al.~2007).
In this case, an alternative to explain the radial concentration is biasing owing to their early formation in rare peaks.
Moore et al.~(2006) showed that the radial distribution of dark matter that collapsed into halos by $z\sim12$, with mass scales $\gtrsim2.5\sigma(M,z)$, is sufficiently biased to match the radial distribution of the Galaxy's BGCs and halo stars.  
A few simulations have tried to capture this early formation in dwarf galaxies.
Bromm \& Clarke (2002)  identified gas clumps  with masses $\sim 10^6 M_{\odot}$ in their simulation
of a proto-dwarf galaxies ($\sim 10^8 M_{\odot}$) as possible proto-GCs, while Kravtsov \& Gnedin (2005) used a 
a subgrid model to split ``cluster particles" in dwarf galaxies with  $M>10^9 M_{\odot}$.  However, these simulations have
not yet demonstrated that GCs do form, nor why a $10^6 M_{\odot}$ cloud should form a GC at high $z$, while such clouds form associations or open clusters today.  

Halo stars are also old,  but their connection with BGC formation remains uncertain.  These stars have abundances similar to BGCs and a consistent radial distribution (e.g., Brodie \& Strader 2006; Helmi 2008).     In the Galaxy, the cluster mortality rate is high (Lada \& Lada 2003; Bastian \& Goodwin 2006), and the destruction of BGC-like star clusters that formed could produce a substantial fraction, even all, of the halo.  Fall \& Zhang (2001) showed that a power law initial cluster luminosity function (ICLF), like what is seen in the Antennae (Schweizer 1987; Zhang \& Fall 1999), could be shaped into the present-day globular cluster luminosity function (GCLF). 

We focus on BGC formation using arguments that are independent of the details of hydrodynamics and star formation. We show that halo stars and BGCs could have a common origin; we place constraints on the epoch of halo star and BGC formation; and, we argue that simulations of a biased region of the universe, with box sizes $\sim1$ Mpc comoving, should capture BGC formation by $z\sim10$, regardless of the mechanism.  In \S 2, we estimate the mass of the halo star population in the Milky Way Galaxy (MWG) by combining the GCLF with an ICLF.  We then find the maximum redshift that BGC and halo star formation could have been completed in \S 3.  In \S 4, we use radial biasing to constrain star formation efficiencies. We present in \S 5 a proof-of-concept simulation of the formation of star clusters at high $z$.  Our conclusions are summarized in \S 6.

\section{Luminosity Functions and Mass Estimates}

In this section, we calculate the total initial mass of MWG halo stars and BGCs if they formed with an Antennae-like initial cluster mass function (ICMF).  
The GCLF is $\Phi\propto \exp(-(V-V_0)^2/2\sigma_V^2)$, for absolute visual magnitude $V$ and  turnover magnitude $V_0$.  The globular cluster mass function (GCMF) can be estimated by assuming a constant mass-to-light ratio $\Upsilon_V$: $f_{\rm GC}\propto \exp(-(\mu-\mu_0)^2/2\sigma_\mu^2)$, where $\mu=\log m$ for mass $m$. We set $V_0=-7.3$ and $\sigma_V=1.2$, which is consistent with the Harris (1991) and Gnedin (1997) fits for MWG GCs.  Assuming 
$\Upsilon_V=2$ gives $\mu_0=5.15$ and $\sigma_{\mu}=0.48$. 
We normalize the mass function to the Milky Way's current BGC population adopting a total number $N_{GC}\sim100$.   To represent the ICMF, we follow Zhang \& Fall (1999) with an ICMF $f_{SC}= m dN/dm \propto m^{-1}$.  Figure 1 shows curves for $f_{GC}$ and $f_{SC}$, where $f_{SC}$ is normalized such that it osculates $f_{GC}$ at $\mu_{\rm osc}=\mu_0+2.3 \sigma_{\mu}^2$. Integrating $f_{SC}$ yields a total mass of $104 m_{\rm osc} \left( \mu_{\rm high}-\mu_{\rm low}\right)$. Including star clusters between -2 and -12 $V$, the total mass in the initial power law distribution is  $2\times10^8 M_{\odot}$, compared with $\sim2\times10^7 M_{\odot}$ for the present-day BGCs.    

This calculation underestimates the ICMF as even clusters with masses of $m_{\rm osc}$ will suffer destruction.
 Typical GC destruction rates are 0.03/Gyr are 0.1/Gyr for evaporation and evaporation+disk-shocking, respectively
(Gnedin \& Ostriker 1997).    Assuming these are constant over 13 Gyr,  the mass function is reduced by a factor of 1.5 or 4.
Therefore the mass in halo stars that could be produced during BGC formation is between 3 and $8\times 10^{8} M_{\odot}$, consistent with estimates of the old stellar halo mass (e.g., Binney \& Tremaine 2008).

Other authors have investigated whether the halo population could be produced by dissolved clusters.  Surdin (1995) 
concluded that cluster dissolution could not produce the mass  the entire spheroid  (bulge+halo; $\sim5\times 10^{9} M_{\odot}$)
(see also Ostriker \& Gnedin 1997).   
However, the bulge population is chemically different from the stellar halo and the BGCs.  
The peak of the metallicity distribution in the bulge is [Fe/H] $\sim -0.1$ and extends to $\sim -1$ (Zoccali et al.~2003), while the halo and BGCs have peaks at [Fe/H] $\sim -1.6$ and extend to $\sim -3$ (Helmi 2008, Zinn 1985). 
Present thoughts on the bulge are that it is a pseudo-bulge formed through disk processes (see Binney 2009 for a review) and should be left out of this comparison.

Henceforth, we assume that  the {\it halo star} population 
came from dissolved clusters that formed along with the
present-day BGCs.  
It is unclear whether the  Antennae-like ICMF describes 
star clusters at high $z$, but the detailed shape of the ICMF is
not central to our argument.   We only require that a distribution of
clusters is produced along with the BGCs, and an Antennae ICMF
gives us a fiducial model. We find a
minimal extrapolated mass of $2\times10^8 M_{\odot}$.  Including
cluster evaporation, the mass in halo stars produced
during BGC formation is consistent with 3-$8\times10^8 M_{\odot}$.
Hereafter, we assume this mass to be $M_h\sim5\times10^8 M_{\odot}$,
consistent with observational constraints on the MWG's
stellar halo.

\section{Required Star Formation Efficiency}

We now estimate the maximum redshift that BGC formation could have been completed 
by requiring enough baryons in appropriate collapsed objects to form the old stellar halo.
To do so, we use the Sheth \& Tormen (2002) formalism for
the conditional mass function.
We set the critical density for collapse $\delta_c=1.686$. The value for a density fluctuation $\delta$ for a given $z$ is determined by scaling $\delta_c$ with the growth function.  The mass variance is calculated using a top-hat filter in real space. We use the Eisenstein \& Hu (1998) transfer function with baryonic acoustic oscillations.  
The conditional mass function is integrated to find the mass in halos with $M\ge M_{\rm min}$ at $z$ 
in a halo of mass  $M_0$ at $z_0$. 

We take $M_0=10^{12} M_{\odot}$ at $z_0=0$, and assume WMAP5 cosmology ($\Omega_c=0.214$, $\Omega_b=0.044$, $\Omega_{\Lambda}=0.742$, and $h=0.719$;  Hinshaw et al.~2009).  The slope of the power spectrum $n=0.95$, and $\sigma_8=0.8$. The minimum mass for a halo $M_{\rm min}$ is set by the lowest mass we expect to be relevant to BGC formation. This mass must be $\gtrsim 10^6 M_{\odot}$, i.e., where we expect the first stars to form (e.g., Bromm et al.~2009).  Figure 2 shows the cumulative baryon mass in collapsed objects, assuming the universal baryon fraction, for minimum dark matter halo masses of  $10^6$, $10^7$, $10^8$, and $10^9 M_{\odot}$.  The two thin horizontal lines show the mass fraction in baryons required to form $\sim5\times10^8 M_{\odot}$ of halo stars and BGCs, assuming a global star formation efficiency (SFE) of $\eta=1$\% (top) and 10\% (bottom).
Reionization may well be responsible for the truncation of BGC formation (see Brodie \& Strader 2006), hence we show the WMAP5 reionization redshift with gray shading in Figure 2.  The plot shows that halo star and BGC formation could have been completed before reionization for SFEs $\sim10$\%, if BGC formation took place in halos less massive than $10^8 M_{\odot}$.  This does not mean that BGCs could not form in higher mass halos; instead, it means that they could not form exclusively in high-mass halos.  Although reionization could be responsible for halting BGC formation after $z\sim10$, reionization is not central to our argument, as we shall see next.
 
\section{Radial Biasing Constraints}

Diemand et al.~(2005) and Moore et al.~(2006) used dark matter only simulations to demonstrate  that material that collapses into halos with mass scales of $\nu\gtrsim2.5$ at high redshift, where  $\nu=\delta_c/\sigma(M,z)$, has a radial distribution in the $z=0$ galaxy that is consistent with the MWG halo stars' and BGCs'.  We note that the characteristic mass, i.e., the knee in the Press-Schechter function, corresponds to $\nu=\sqrt{6}$ (Binney \& Tremaine 2008). 
Using the Sheth-Tormen conditional mass function, we integrate over all halos with masses greater than $\sigma(M,z)$ for $\nu=2.5$.  At $z\sim12$, this mass fraction $f_{\nu\ge2.5}\sim0.027$ (thick solid line in Figure 2).  
To produce the halo and BGC population from exclusively high-$\nu$ halos, the global SFE needs to be $\gtrsim 10$\%. We analyzed the Diemand et al.~simulation data for comparison with the Sheth-Tormen formalism.  We reproduce the biasing results, so the radial distributions of halo stars and BGCs are consistent with $\nu=2.5$ halos from $z>10$.  However, more collapsed objects were found in the simulations than predicted by the conditional mass function for $z\gtrsim 5$.  The biased mass calculated from Sheth-Tormen may be a lower limit. 

For a 10\% SFE, about 10\% of halo stars and BGCs could be produced by $z\sim18$.  For the same SFE, the rest of the stars can be produced by $z\sim13$, 120 Myr later. The Lagrangian volume for the MWG is approximately 27 Mpc$^3$ comoving.  Using these figures, we estimate the star formation rate required to form $5\times10^8 M_{\odot}$ of stars is $\sim0.15 M_{\odot}\rm ~yr^{-1}~Mpc^{-3}$. This value is similar to the observed peak in the universal star formation rate (Madau et al.~1998; 0.12-0.17 $M_{\odot}\rm ~yr^{-1}~Mpc^{-3}$ at $z\sim1.5$).  If BGC formation extends to $z\sim10$, our estimate would be about a factor of two lower.  Note that the universe is only about 500 Myr old by $z\sim10$;  de Angeli et al.~(2005) found an age spread in their low-metallicity GC sample that is less than 600 Myr and consistent with zero.  

Combining our above arguments, we conclude that capturing the formation of halo stars and BGCs in $\Lambda$CDM using numerical techniques requires only a small, biased box.  BGCs  should form in a $\sim 1 $ Mpc comoving box by $z\sim10$, regardless of the underlying process.   Based on the Sheth-Tormen mass function, the simulation volume should contain three $10^8 M_{\odot}$ halos by the end redshift.   If BGCs do not form in such a simulation, there is likely a problem with the subgrid model or the underlying cosmology.  

\section{Test Simulation}

\subsection{Setup}
We highlight preliminary simulation results that will be presented in detail in a forthcoming paper.  We employ RAMSES (Teyssier 2002), an adaptive mesh refinement cosmology code that solves the equations of hydrodynamics using a second-order unsplit Godunov method and integrates particles using particle-mesh techniques.  The code has been augmented to follow nonequilibrium chemistry for a nine-species gas: e, HI, HII, HeI, HeII, HeIII, H$^{\rm -}$, H$_2$, and H$_2^+$.   Cooling due to collisions between H$_2$ and the other species is modeled according to Glover \& Abel (2008).  
The Dubois \& Teyssier (2008) star formation recipe is modified to capture pop III star formation and supernova feedback.  Wherever the gas number density is greater than the critical value $n_0=10^5 \rm cm^{-3}$, there is a chance to form a star particle.
The local SFE is set to 1\% per dynamical time.  

If the metallicity of the gas is above $Z_{\rm thresh}=5\times10^{-7} Z_{\odot}$, the resulting star particle represents a distribution of population II stars.  This threshold is lower than suggested by Bromm \& Loeb (2003) because we wanted to explore whether very low-metallicity population II stars could form if given the opportunity.  Ten percent of the star particle's mass is injected back into medium after 10 Myr as supernova ejecta, with 0.1 of the ejecta mass in metals.  Below $Z_{\rm thresh}$, the particles represent a distribution of population III stars.  

The IMF for population III stars remains unclear. The minimum mass may be $\gtrsim 10 M_{\odot}$ (e.g, Bromm et al.~2009), but this remains highly uncertain.  Likewise, we know neither the maximum mass for a pop III star  nor 
the importance of pair-instability supernovae to the enrichment of the second generation of stars.  Tumlinson et al.~(2004) argue that the observed element ratios of old pop II stars are inconsistent with enrichment by pair-instability supernovae.  Moreover, the IMF shows little sensitivity to metallicity (Kroupa 2001). Whether this extends to extremely low metallicities remains unknown. In the absence of better constraints on pop III star formation, we assume a Salpeter IMF ($dN/dM\sim M^{-2.35}$) with a low-mass cutoff $M_{\rm co}$ (see below). Whenever a star particle is formed, the particle's mass $M_{\rm sp}$ is distributed among a random distribution of stars.  The probability that a star is between masses $M_0$ and $M_1$ is given by $P=A(M_1^{-1.35}-M_0^{-1.35})$, where $A=(M_{\rm top}^{-1.35}-M_{\rm co}^{-1.35})^{-1}$ and $M_{\rm top}=1000 M_{\odot}$ is the most massive star in the distribution.   A given star mass is selected by setting $M_{\rm star}=(r/A+M_{\rm co}^{-1.35})^{-1/1.35}$, where $r$ is a random variable with an even distribution between 0 and 1.  This process continues until the total mass in stars is equal to $M_{\rm sp}$.   We set  $M_{\rm co}=3 M_{\odot}$, which is based on the theoretical calculations of Nakamura \& Umemura (2001), who find that the smallest Jeans mass for which the gas becomes opaque to H$_2$ lines is $\sim 1 M_{\odot}$.

For each population III star in the distribution, its contribution to ejecta, metal enrichment, long-lived stars and remnants, pair-instability supernovae, and black holes is tracked (see Table 1).  The remnant neutron star or black hole mass is determined, when applicable, by $M_{\rm R/BH}={\rm MAX}( M_{\rm star} 0.1^{13.5 M_{\odot}/M_{\rm star}},\ 1.35 M_{\odot} )$. This prescription roughly follows the results of Timmes et al.~(1996) for population III stellar remnants.  Metal enrichment is included using $Z=0.4(M_{\rm star}-13.5 M_{\odot})/M_{\rm ejecta}$ (Woosley \& Weaver 1982).  Whenever a supernova occurs, for either population II or III, $M_{\rm ejecta} 10^{50} {\rm erg}~M_{\odot}^{-1}$ is deposited directly into the cell containing the star. 

We use a box size of 589 $h^{-1}$ kpc, and the cosmology is set to $h=0.719$, $\Omega_{\rm matter}=0.258$, $\Omega_{\Lambda}=0.742$, and  $\Omega_b=0.045$. The power spectrum normalization is set to $\sigma_8=1.0$.  The box size is smaller than recommended above, but it is intended to be a proof-of-concept simulation. The highest refinement at a given expansion factor $a$ is kept at a physical resolution $\Delta x_{\rm min}\sim 0.5 $pc, which corresponds to refinement level $l=17$ at $z\sim13$.  For dark matter particles only, the highest level of refinement used in the cloud-in-cell approximation is $l=13$ to avoid two-body relaxation effects (e.g., Knebe et al.~2000; Levine et al.~2008).  The initial conditions were generated using the COSMICS package (Bertschinger 1995).  The simulation box has a 256 cubed region (236 h$^{-1}$ kpc) nested within a coarser 128 cubed grid. The dark matter particle mass in the higher resolution region is about 1000 $M_{\odot}$.

\subsection{Simulation Results}

Figure 3 shows maps of the average number density and temperature along the line of sight in a 360 pc cubed region at $z\sim13.3$.  The region contains a $5.3\times10^6 M_{\odot}$ dark matter halo with about $10^6 M_{\odot}$ of gas within $R_{200}\sim 420$ pc.  Two star clusters are present in this halo, with a combined mass $\sim 9\times10^4 M_{\odot}$.  Most of this mass is in pop II stars, but about $138M_{\odot}$ is from the pop III generation, either trapped in neutron stars or in stars that will become white dwarfs.  Although pop III black holes were not formed in these clusters, but some clusters in the simulation have a total black hole mass  $\gtrsim 1000 M_{\odot}$.  More than half of the combined mass occurred in a burst of star formation that lasted about $\sim30$ Myr. The effects of supernovae on the surrounding gas can be seen in the temperature and density plots, where bubbles have formed.   The initial enrichment for the stars in Figure 3 was set by population III stars with a combined mass of $\sim190 M_{\odot}$. The final metallicity distribution is shown in the bottom-left panel, with a peak $\rm [Fe/H]\sim-1.3$, consistent with Morgan \& Lake (1989).  Greif et al.~(2007) found that a 200 M$_{\odot}$ population III star, which did not occur in our simulation, could expel the gas from a halo, but that owes to the hard radiation from the precursor star and the extreme injection of $10^{52}$ erg.   The simulation demonstrates that the formation of star clusters and their subsequent evolution can be captured from cosmological initial conditions. If these clusters survive to be called BGCs today, they represent a mode of GC formation that takes place in dark matter minihalos.  This formation channel could operate until reionization limits cooling (Bromm \& Clarke 2002).  Cluster formation may then be delayed until massive disk galaxies form, which may be responsible for the red GC population.

\section{Conclusions}

During BGC formation,  $3\times10^8$ to $8\times 10^8 M_{\odot}$ of stars should have been produced.  A large fraction, if not all, of the halo stars are remnants of destroyed BGCs and lower mass clusters that formed with BGCs.  Using the conditional mass function, we found that the entire halo population could be produced by $z\sim 13$ with star formation efficiencies of $\sim10$\%.   Based on constraints from reionization, radial biasing, and Press-Schechter theory, we argue that a simulation of a biased, 1 Mpc comoving region of the universe should capture the formation of BGCs by $z\sim10$.  If such a simulation does not capture the formation of these systems, it most likely reflects a problem in the subgrid model or cosmology.  Finally, we presented a simulation that shows the formation of BGC-like clusters with masses $\sim 5\times10^4 M_{\odot}$ by $z\sim13$.   These clusters form in dark matter minihalos and can self-enrich to metallicities that are consistent with observed BGCs.

This research was supported by an SNF Grant and the Zurich CTS.  Simulations were run on the zbox machines, maintained by ITP UZH.  We thank the anonymous referee for comments that improved this manuscript.  We thank Ben Moore and Joachim Stadel for useful discussions and for making the Diemand et al.~data available.

\begin{table}
\begin{center}
\begin{tabular}{ lll }\hline
Mass Range $M_{\odot}$ & Outcome & $t_{\rm SN}$ (Myr)   \\\hline
$<8$ & $M_{LL}$ & --\\
$8<M<25 $ & $M_{LL}$, $M_{\rm ejecta}$ & 10\\
$25<M<35 $ & $M_{BH}$, $M_{\rm ejecta}$ & 10 \\
$35<M<140 $ & $M_{BH}$ & -- \\
$140<M<260 $ & $M_{\rm ejecta}$ & 3\\
$260<M $ & $M_{BH}$ & --\\\hline
\end{tabular}
\caption{The mass in long-lived stars and remnants, $M_{LL}$, black holes, $M_{BH}$, and ejecta mass $M_{\rm ejecta}$.  The remnant mass is determined by $M_{\rm R/BH}={\rm MAX}( M_{\rm star} 0.1^{13.5 M_{\odot}/M_{\rm star}},\ 1.35 M_{\odot} )$, which is based on the results of Timmes et al.~(1996). The ejecta mass is the difference between the star's mass and its remnant.  In the case of a pair-instability supernova ($140 < M<260 M_{\odot}$), the ejecta mass is the star's mass. The third column is the supernova delay time. }
\end{center}
\end{table}

\begin{figure}[h]
\begin{center}
\includegraphics[width=12cm,angle=270]{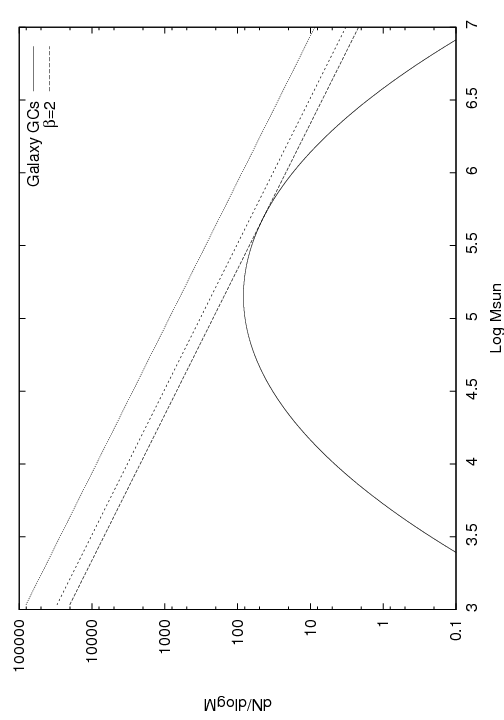}
\caption{The solid curve represents the MWG  GCMF, as seen today, normalized by 100 clusters.  The lines represent the ICMF  $mdN/dm\propto m^{-1}$.  The lowest line shows the osculation mass $m_{\rm osc}$, the dashed line accounts for evaporation, and the dotted line includes evaporation+disk-shocking. }
\end{center}
\end{figure}

\begin{figure}[h]
\begin{center}
\includegraphics[width=12cm, angle=270]{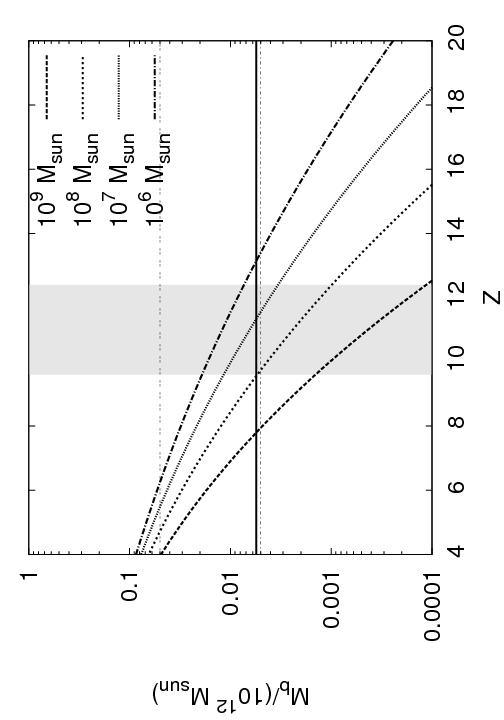}
\caption{Conditional mass function for a halo of $10^{12} M_{\odot}$ at $z=0$.  The curves represent the baryon fraction in collapsed objects for different dark matter halo minimum mass cutoffs.  The thin, horizontal lines show the baryonic mass that is required to produce $5\times10^8 M_{\odot}$ of halo stars and BGCs for a global star formation efficiency of 1\% (top) and 10\% (bottom). The thick line shows the fraction of mass that comes from $\nu \gtrsim \nu_{2.5}$ (\S 4).  The gray region shows the WMAP5 measurement for reionization.}
\end{center}
\end{figure}

\begin{figure}[h]
\begin{center}
\includegraphics[width=12cm]{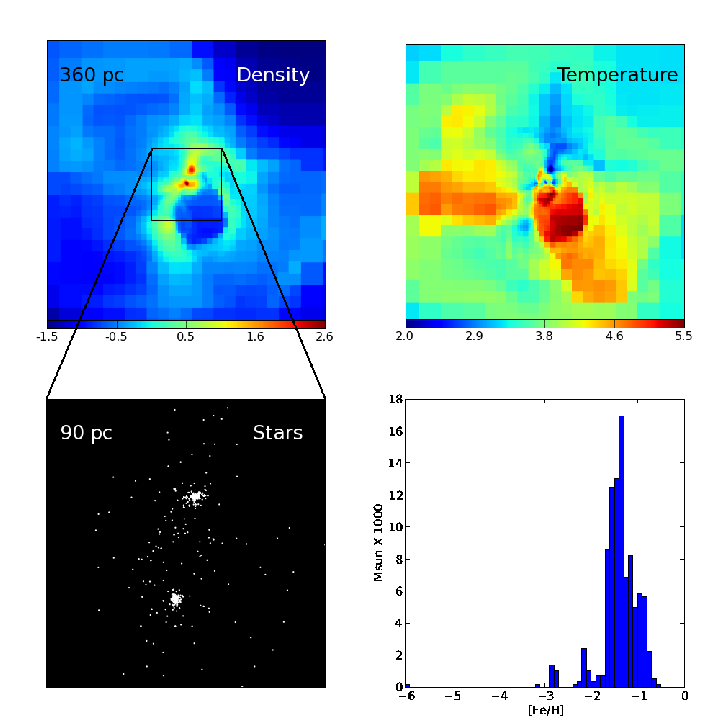}
\caption{Snapshot from a RAMSES simulation with nonequilibrium chemistry and the star formation prescription described in \S 5.  The images correspond to $z\sim 13.3$.  Top-left: average number density ($\rm cm^{-3}$) along the line-of-sight.  Top-right: average density-weighted temperature over the mean molecular weight ($T/\mu~\rm K$).  Bottom-left: star clusters in a dark matter halo $\sim 5\times10^6 M_{\odot}$.  Bottom-right:  histogram of stellar metallicities for the stars shown in the bottom-left panel. }
\end{center}
\end{figure}

\end{document}